\documentclass[twocolumn]{aastex62}

\pdfoutput=1
\accepted{December 17, 2018}

\shorttitle{UV/X-ray Comptonization delay in Mrk~493}
\shortauthors{Adegoke et al.}

\begin{document}

\title{UV to X-ray Comptonization delay in Mrk~493}

\correspondingauthor{Oluwashina Adegoke}
\email{oluwashinaa@iisc.ac.in}

\author{Oluwashina Adegoke}
\affil{Department of Physics, Indian Institute of Science, Bangalore 560012, India}

\author{Gulab C. Dewangan}
\affiliation{Inter-University Center for Astronomy and Astrophysics, Pune 411007, India}

\author{Pramod Pawar}
\affiliation{Inter-University Center for Astronomy and Astrophysics, Pune 411007, India}
\affiliation{Swami Ramanand Teerth Marathwada University, Nanded 431606, India}

\author{Main Pal}
\affiliation{Centre for Theoretical Physics, Jamia Millia Islamia, New Delhi 110025, India}



\begin{abstract}

 The broadband X-ray emission from type 1 active galactic nuclei, dominated by a powerlaw continuum, is thought to arise from repeated inverse Compton scattering of seed optical/UV photons by energetic electrons in a hot corona. The seed optical/UV photons are assumed to arise from an accretion disc but a direct observational evidence has remained elusive.  
Here we report the discovery of variations in the UV emission preceding the variations in the X-ray emission based on $\sim100\,\mathrm{ks}$ \textit{XMM-Newton} observations of the narrow-line Seyfert 1 galaxy \rm{Mrk~493}. We find the UV emission to lead by $\sim5\,\mathrm{ks}$ relative to the  X-ray emission. The UV lead is consistent with the time taken by the UV photons to travel from the location of their origin in the accretion disc to the hot corona and the time required for repeated inverse Compton scattering converting the UV photons into X-ray photons. Our findings provide first direct observational evidence for the accretion disc to be responsible for the seed photons for thermal Comptonization in the hot corona, and constrain the size of the corona to be $\sim20r_{g}$.
\end{abstract}



\section{Introduction} \label{sec:intro}

A substantial fraction of the emission from active galactic nuclei (AGN) arises from the accretion disc in the optical/UV bands and from the hot, relativistic particles constituting the ``corona''. The disc emission is of thermal nature and can be approximated as a multi-color blackbody emission in the optical/UV band \citep{1999PASP..111....1K} while the coronal emission can extend to hard X-rays in the form of non--thermal powerlaw component with a high energy cut-off \citep{1993ApJ...413..507H, 2015MNRAS.451.4375F}.

The emission from AGN vary strongly on a wide range of timescales and over the entire electromagnetic spectrum \citep[see e.g.,][]{
2003ApJ...593...96M, 2004PThPS.155..170U, 2010MNRAS.403..605B},
 however, the mechanism that drives this variability and in particular the inter-band correlations is still a subject of active research.

There has been remarkable progress in proffering explanations to this exotic variability behavior. For example, \citet{1991ApJ...371..541K} argued that changes in the X-ray continuum properties which illuminates and heats up the disc causes the optical/UV continuum to vary. This implies naturally that the optical/UV emission should lag the illuminating X-rays in their variability \citep{2007MNRAS.380..669C}. This has been observed in several sources \citep[see e.g.,][]{2007MNRAS.380..669C, 2009MNRAS.397.2004A, 2012MNRAS.422..902C,
2016MNRAS.456.4040T, 2017MNRAS.466.1777P, 2018MNRAS.tmp.1900M}. However, as pointed out by \citet{2007ASPC..373..596G}, reprocessing alone can be ruled out as the dominant mechanism of variability in many AGN through simple energetics argument. This is because the ``big blue bump" dominating the bolometric luminosity exceeds significantly the X-ray luminosity responsible for reprocessing. A few sources have shown a correlation consistent with zero or no lag, some do not reveal any correlation and others show more complex variability pattern \citep{2002AJ....124.1988M, 2008MNRAS.389.1479A, 2009MNRAS.394..427B, 
2017MNRAS.472.2823P, 2018MNRAS.475.2306B}. In the specific cases of NGC~5548 as well as NGC~4151, the X-ray to UV/optical relationship is complex and difficult to explain as solely due to reprocessing \citep{2015ApJ...806..129E, 2017MNRAS.470.3591G, 2017ApJ...840...41E}. With respect to the inward propagation model -- an alternative model -- the optical/UV emission may lead the X-rays since the optical/UV photons emanate further out from the central engine. While propagation delays on viscous timescales are yet to be confirmed, possible combination of X-ray reprocessing and propagation fluctuations on different timescales can explain the observed X-ray/optical correlations in some AGN \citep[see e.g.,][]{2005A&A...430..435A, 2013MNRAS.433.1709G, 2003MNRAS.343.1341S}.

Although it has not been conclusively established so far, Compton upscattering of optical/UV seed photons into X-rays in the hot corona can provide a compelling explanation to optical/UV/X-ray correlated variability seen in AGN. In such a case, the thermal optical/UV seed photons drive changes in the X-ray emission. This will imply that the optical/UV seed photons lead the X-rays in their variability by the sum of light crossing and the Comptonization timescales.

The most suitable AGN to probe the Comptonization delay are those with low black hole masses e.g., the least massive narrow-line Seyfert 1 (NLS1) galaxies.
\rm{Mrk~493} is one such NLS1 known for its uniquely strong $\rm{Fe\,II}$ emission \citep{1985ApJ...297..166O}. 
From their reverberation mapping campaign, \citet{2014ApJ...793..108W} measured the mass of $\rm{Mrk\,493}$ to be $\sim1.5\times10^{6}\,M_{\odot}$.
We study temporal characteristics of Mrk~493 using the long \textit{XMM-Newton} observation of Mrk~493 performed in 2015 \citep{2018MNRAS.tmp..812B}. 
This paper is structured as follows. In Section 2, we describe the observation and data reduction procedure. Section 3 focuses on the data analysis and result. In Section 4, we discuss the implication of our result and conclude.

\section{Observations and Data Reduction}
The \textit{XMM-Newton} satellite \citep{2001A&A...365L...1J} observed \rm{Mrk~493} twice in 2015, first on Feb. 24 (observation ID 0744290201) and the second on Mar. 2 (observation ID 0744290101) each for a duration of $\sim100\,\mathrm{ks}$. The observations were carried out with all European Photon Imaging Camera (EPIC)
\citep{2001A&A...365L..18S, 
2001A&A...365L..27T}, the Reflection Grating Spectrometer (RGS) and the Optical Monitor \citep[OM;][]{2001A&A...365L..36M}. Data from the second observation (0744290101) was not used in this study because OM was operated only in the \textit{image} mode, also several filters were used which reduces considerably the staring time for each of the used filters.   

\begin{figure}
\centering
\includegraphics[width=.50\textwidth, height=0.25\textheight, angle=0]{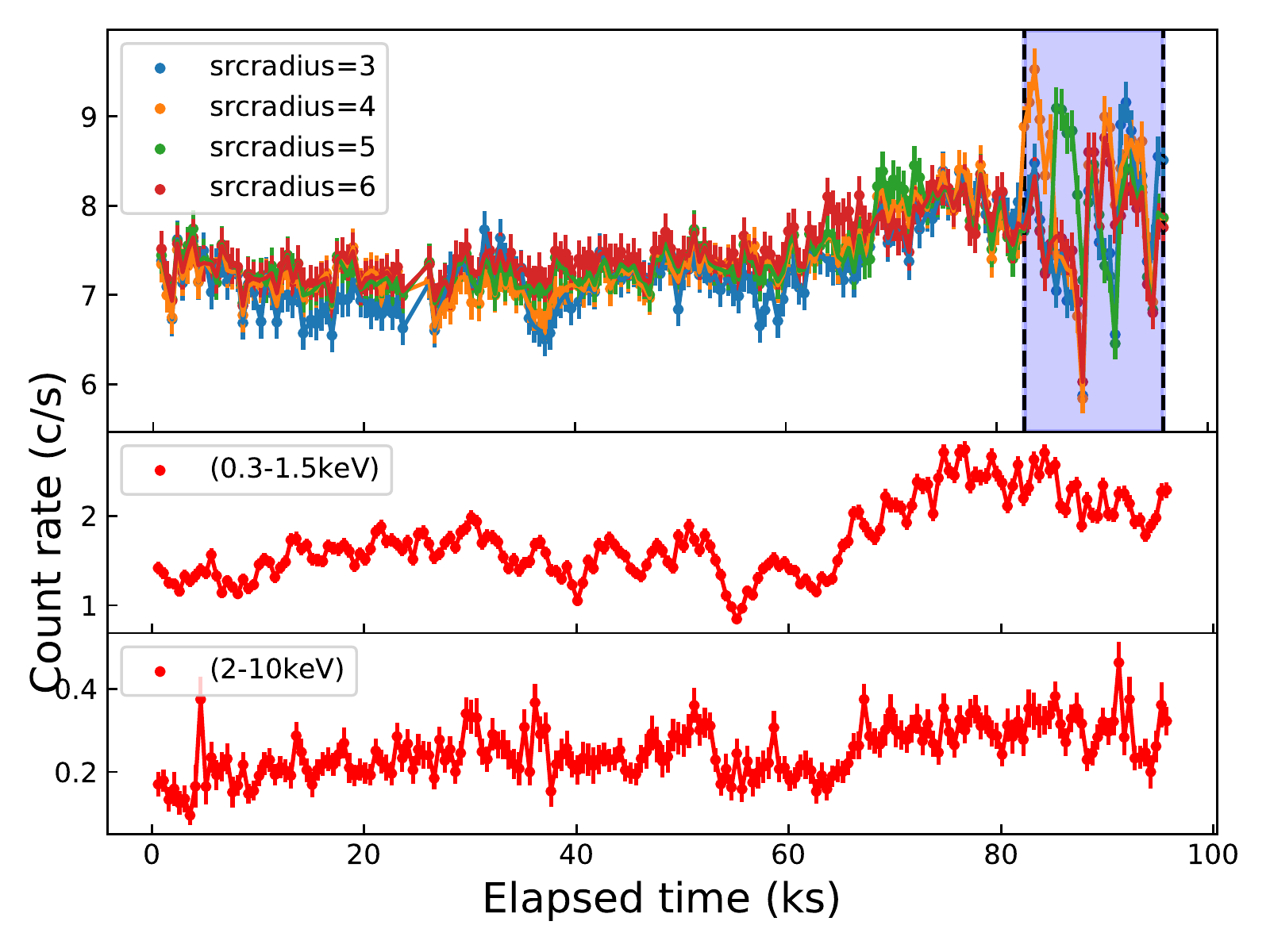}  
\caption{The UVW1, SX and HX lightcurves of \rm{Mrk 493}. The UVW1 lightcurves were extracted with 4 different source radii as indicated in the panel, the shaded blue region in the uppermost panel marks the interval containing the UVW1 frames where the source offset correction appears to be unreliable as explained in Section 2.}
	\label{obs2lc}
\end{figure}

We employed the Science Analysis System (SAS v.16.1.0) package for data reduction with updated Current Calibration Files (CCFs). We generated event files for the pn and MOS detectors. We extracted the event file list using the task EVSELECT. The datasets were then screened individually for intervals of high particle background (i.e. flaring) in the lightcurve to produce good time interval (GTI) files which were then used to obtain cleaned event lists in line with standard procedure. We did not find evidence for significant pile-up in the data. We extracted source photons from a circular region of radius $40''$ centered on the source and the background photons from a source-free region of radius $80''$. We generated the background subtracted lightcurve using the task EPICLCCORR.
We extracted the lightcurve of the source in the soft X-ray (SX: $0.3-1.5\,\mathrm{keV}$) and the hard X-ray (HX: $2-10\,\mathrm{keV}$) bands with $500\,\mathrm{s}$ time bin (shown in Fig. \ref{obs2lc}). We used the EPIC-pn lightcurves for our analysis because of its better sensitivity compared to the MOS. We used the MOS lightcurves only for cross-verification. 

The OM observations were carried out in the \textit{image+fast} mode using the UVW1 filter ($\lambda_{eff}=2910\textup{\AA}$), $25$ short exposures were taken during the observation. The meta-task OMFCHAIN was used to extract the events and to generate the UVW1 lightcurve, again with $500\,\mathrm{s}$ bin size. The lightcurve generated is shown in the uppermost panel of Fig. \ref{obs2lc}. The UVW1 lightcurve shows a small scale variability in the first $\sim80\,\mathrm{ks}$ of observation beyond which (particularly the shaded part of the lightcurve) it shows extreme variability with rapid decline and increase in the count rate which is generally not expected from the accretion disc. To probe whether these variations are due to the source itself or some observational artefact, we manually checked all $25$ OM accompanying images. We noted that for a couple of frames, the source was offset from the center of the detector. The maximum offset is about $25\%$. Although the task \textit{omlbuild}, a part of the meta-task OMFCHAIN accounts for possible missing photons for offset sources using the knowledge of the point spread function (PSF), we decided to cross-verify this especially for the few OM frames that reveal erratic fluctuations (the shaded regions of the OM lightcurve in Fig. \ref{obs2lc}). To do this, we imposed decreasing values of the source radius (in pixel units from 6 to 3) and generated OM lightcurves in each case using the command OMFCHAIN. Since for decreasing source radius, less and less source region is expected to be offset, then the generated lightcurves should overlap (nearly) in principle for all frames.

A close look at the uppermost panel of Fig. \ref{obs2lc} shows that the variations overlap for most of the frames as expected except for $7$ frames where the fluxes show huge variations from one another (the shaded region) although with mostly similar pattern. Thus to avoid any ambiguities, we removed these frames from further analysis. This leaves us with 153 OM usable lightcurve data points.

\section{Temporal Analysis and Results}

Figure \ref{obs2lc} shows the UV, SX and HX lightcurves of \rm{Mrk 493}. As evident from the lightcurves, this AGN is very bright and highly variable. The mean count-rate in the UV, SX and HX bands are $7.48\pm{0.20}\,\mathrm{c/s}$, $1.74\pm{0.08}\,\mathrm{c/s}$ and $0.25\pm{0.03}\,\mathrm{c/s}$ respectively. The minimum to maximum flux ratios for the UV (excluding the shaded region in the lightcurve) is $1.2$ while for the soft X-ray and the hard X-ray emission, the values are $3.3$ and $5.5$ respectively.
To quantify the variability of this source, we calculated the fractional variability amplitude $F_{var}$ \citep{2003MNRAS.345.1271V}, a measure of intrinsic variability in a band. 
$F_{var}$ for the UV, SX and HX bands are $3.1\pm{0.2}\%$, $23.7\pm{0.3}\%$ and $19.6\pm{1.1}\%$, respectively.

\begin{figure*}
\centering
\includegraphics[width=.25\textwidth, height=0.20\textheight, angle=0]{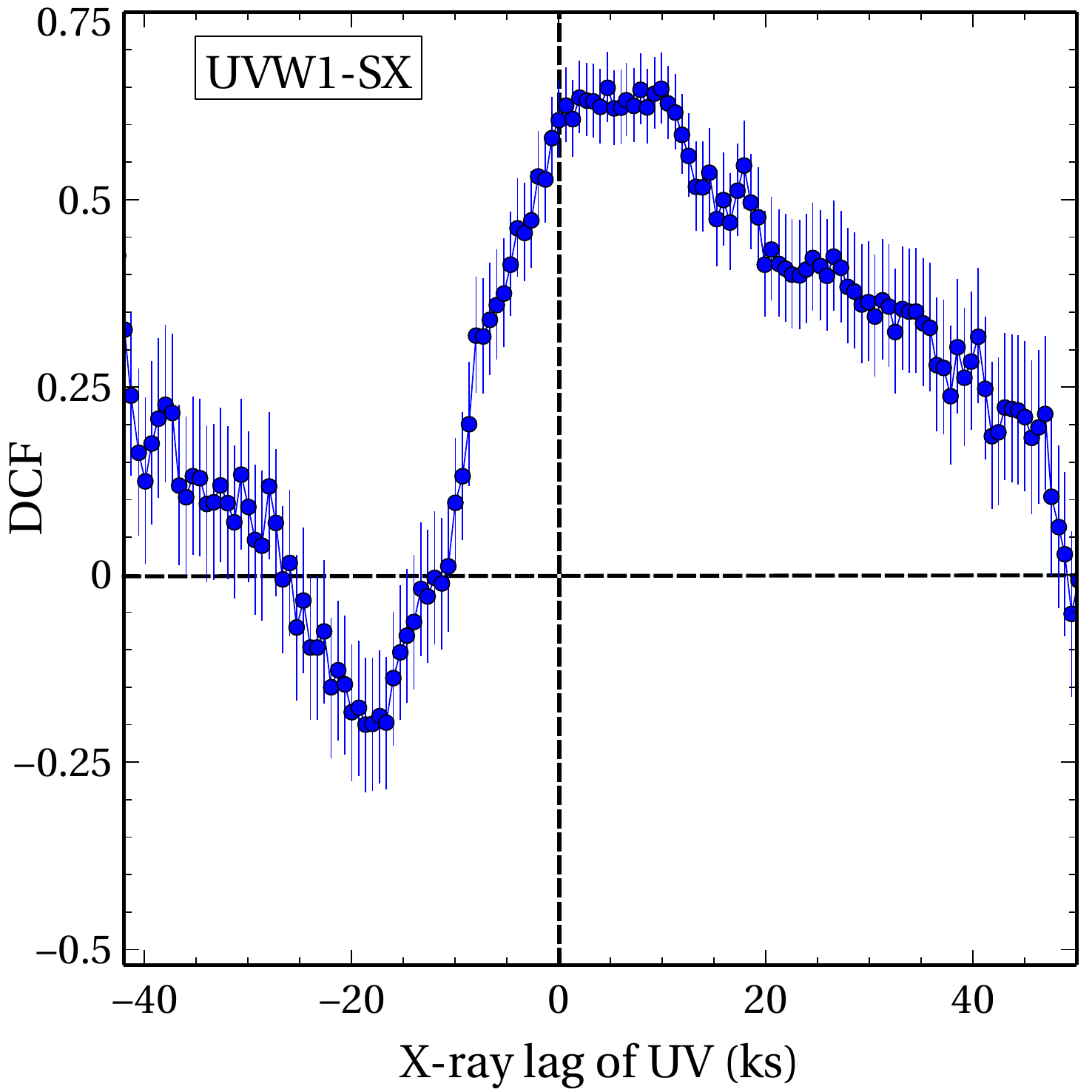}  
\includegraphics[width=.25\textwidth, height=0.20\textheight, angle=0]{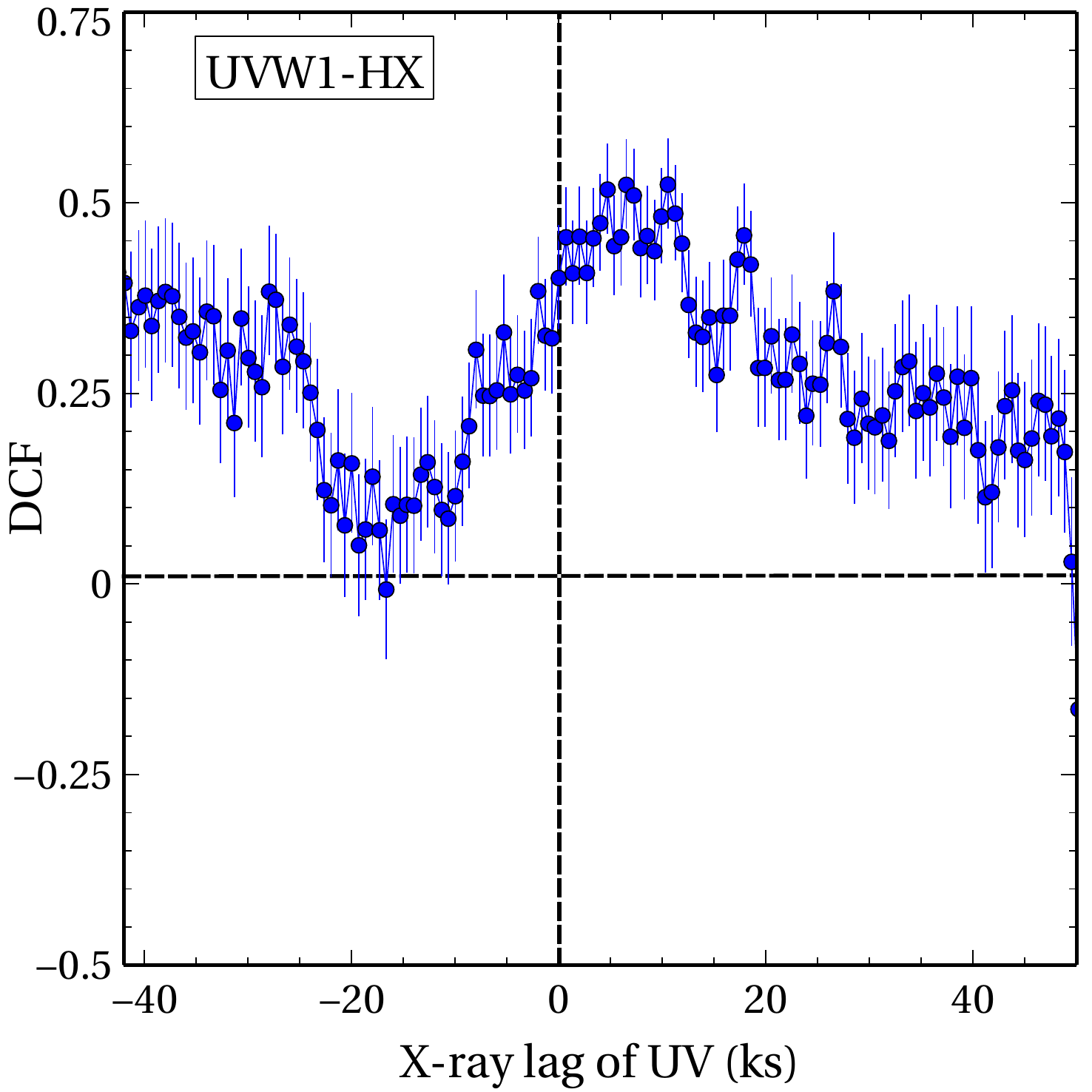} 
\includegraphics[width=.25\textwidth, height=0.20\textheight, angle=0]{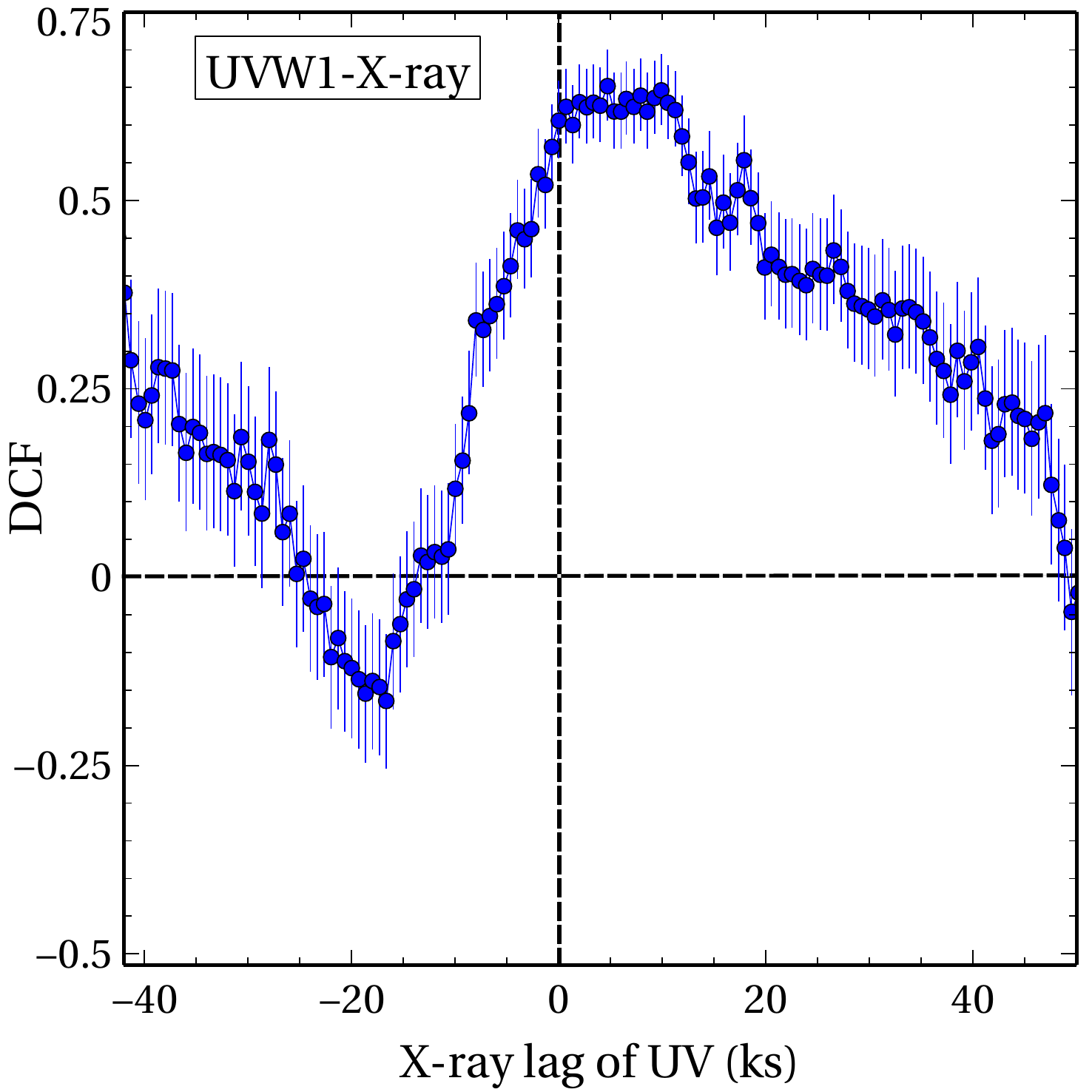}
\caption{\textit{z}-transform based cross-correlation function  between the UVW1 and different X-rays bands: SX, HX, and the full ($0.3-10\,\mathrm{keV}$) band, respectively.}
\label{zdcf}
\end{figure*}

We computed the cross-correlation function (CCF) between the UV and the X-ray lightcurves in order to probe any connection between them. We used the \textit{z}-transformed discrete correlation function (ZDCF) outlined in \citet{1997ASSL..218..163A,2013arXiv1302.1508A}. The ZDCF method uses a variable bin size, keeping at least 11 data points per bin. We imposed 33 data points per bin in our analysis for improved statistics, we further generated 10,000 realizations through the Markov Chain Monte Carlo (MCMC) method to estimate the lags. We have carried out the analysis such that a positive time lag indicates that the variations in the UVW1 band leads. Figure \ref{zdcf} shows the resultant ZDCF observed between the UV and various X-ray bands. The ZDCF reveals a strong and broad peak indicating UVW1 variations are leading the X-ray bands. 
By applying the \textit{p-like} algorithm \citep{2013arXiv1302.1508A} to the output of the ZDCF, we obtained peak likelihood lags for the UV--SX, UV--HX and UV--full X-ray bands. These lags  are given in Table \ref{tab_ccf}.  

\begin{figure}
\centering
\includegraphics[width=.23\textwidth, angle=0]{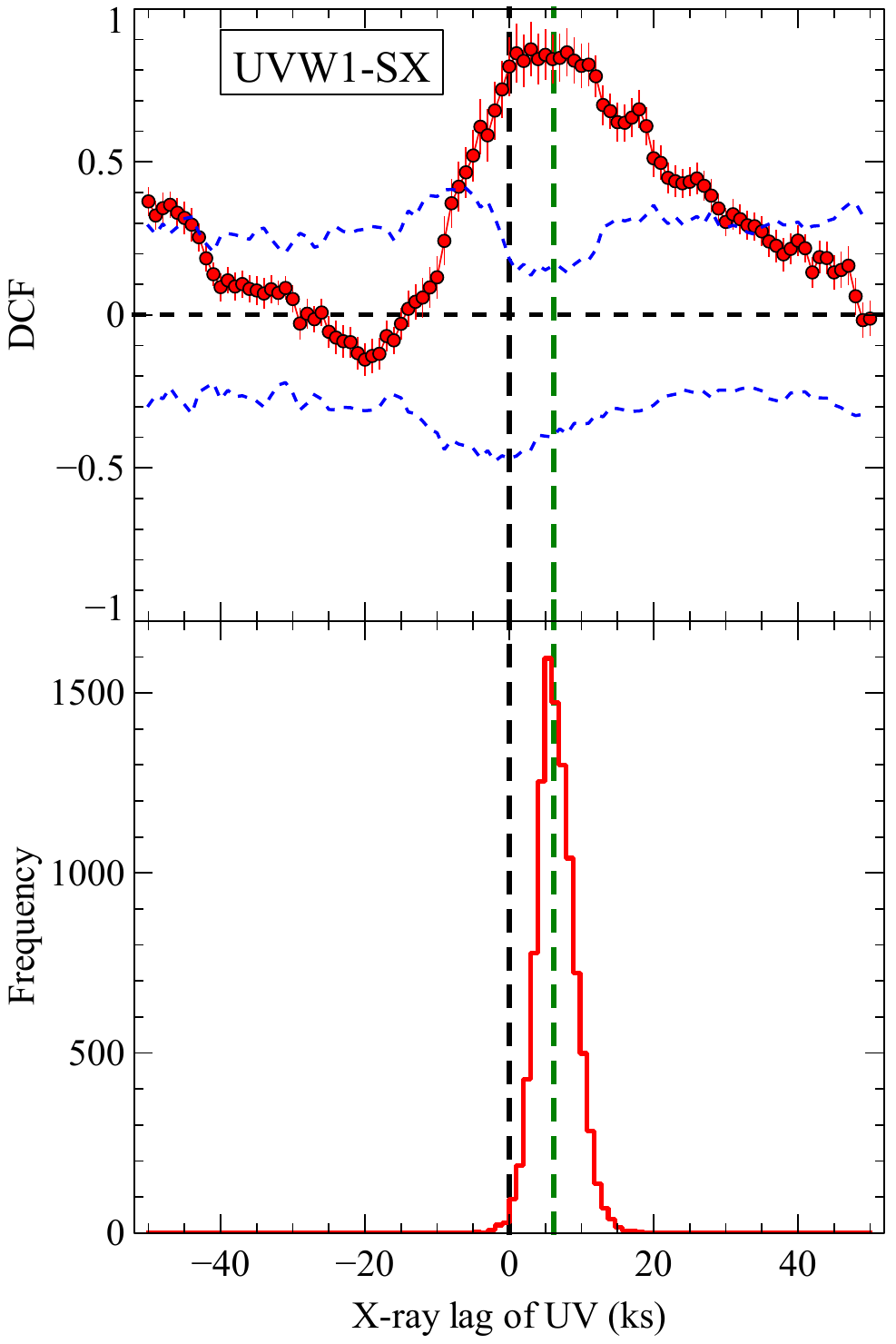}  
\includegraphics[width=.23\textwidth, angle=0]{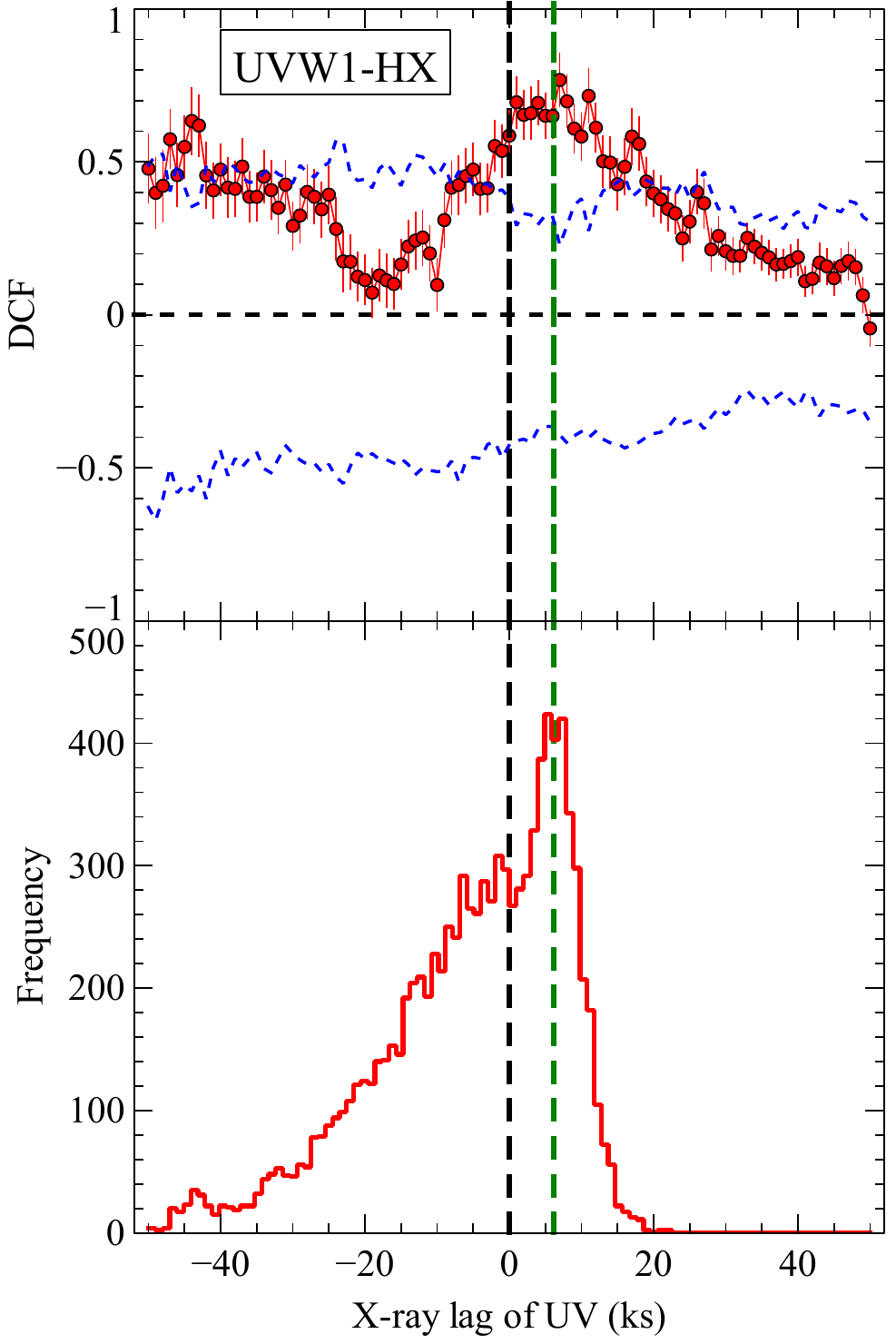} 
\caption{Upper panels: The DCF between the UVW1--SX and the UVW1--HX bands obtained using the method described in \citet{1988ApJ...333..646E}. Bottom panels: Time lag distribution based on 10000 simulated lightcurves using the \textit{bootstrap} technique.}
\label{pydcf}
\end{figure}

To verify the detected positive lag between the UV and X-ray emission, we applied another independent, popularly used technique for estimating lags. We computed cross-correlation function using the Discrete Correlation Function (DCF) of \citet{1988ApJ...333..646E} with \textit{python} implementation (\textit{pydcf}\footnote{https://github.com/astronomerdamo/pydcf}). The DCFs are calculated by using a lag size of $2\,\mathrm{ks}$. We constrained the DCF estimation to the lag range of $\pm40\,\mathrm{ks}$ since the total duration of the observation is $\sim100\,\mathrm{ks}$. As shown in the upper panels of Fig. \ref{pydcf}, a moderately strong correlation is detected between the UVW1 and the X-ray bands with the UVW1 emission leading by $\sim5\,\mathrm{ks}$, consistent with ZDCF. 
In estimating the time lag between two lightcurves, we calculated the mean of all DCF points that are at least $80\%$ of the maximum. We refer to this value as $\rm{DCF_{max}}$ and the corresponding centroid lag mean value as $\tau_{cen}$. This technique gave $\tau_{cen,UV/SX} = 5.5\,\mathrm{ks}$ with $\rm{DCF_{max,UV/SX}} = 0.83$, $\tau_{cen,UV/HX} = 5.2\,\mathrm{ks}$ with $\rm{DCF_{max,UV/HX}} = 0.69$ and $\tau_{cen,UV/X-ray} = 5.5\,\mathrm{ks}$ with $\rm{DCF_{max,UV/X-ray}} = 0.81$ for the UV--SX, UV--HX and the UV--full X-ray bands, respectively.
\begin{table}
\caption{X-ray lags obtained from cross-correlation analysis by different methods}
\centering 
\begin{tabular}{l l l l l l l}
\hline\hline
 & ZDCF ($\mathrm{ks}$) & DCF ($\mathrm{ks}$) & JAVELIN ($\mathrm{ks}$) \\
\hline
UV/SX & $4.7^{+4.8}_{-2.8}$ & $5.5\pm{0.05}$ & $4.7\pm0.3$ \\ 
UV/HX & $10.5^{+1.0}_{-5.6}$ & $7.0^{*}$ & $7.5\pm0.9$ \\ 
UV/X-ray & $4.7^{+4.9}_{-2.7}$ & $5.5\pm{0.04}$ & $4.7\pm0.2$ \\
\hline
\end{tabular}
\textit{Note}: * We chose the mode value of the distribution due to its shape.
\label{tab_ccf}
\end{table}

To estimate the significance of the detected lag, we implemented the Monte Carlo method described in \citet{1998PASP..110..660P}. We created 10,000 pairs of synthetic lightcurves using the technique of Random Subset Selection (RSS) outlined in \citet{1998PASP..110..660P} and then calculated the $\rm{DCF}$ of each pair. 
The dashed blue lines in the upper panels of Fig. \ref{pydcf} show the $90\%$ confidence limit on the estimated lags. Following the same approach as with the observed lightcurves, we computed the centroid lag value for each pair of simulated lightcurves $\tau_{cen,sim}$ and the corresponding $\rm{DCF_{max,sim}}$. With these values, we constructed the sample distribution function of $\tau_{cen, sim}$ i.e. the cross correlation peak distribution (CCPD) of lags. The resulting distributions are shown in the lower panels of Fig. \ref{pydcf} and the measured values (from Gaussian fit) are quoted in Table \ref{tab_ccf}.
The fact that they show good agreement with ZDCF supports our claim that the X-ray emission lag the UVW1 emission by $\sim5\,\mathrm{ks}$.

\begin{figure}
\centering
\begin{tabular}{@{}cc@{}}
\includegraphics[scale=0.21, angle=0]{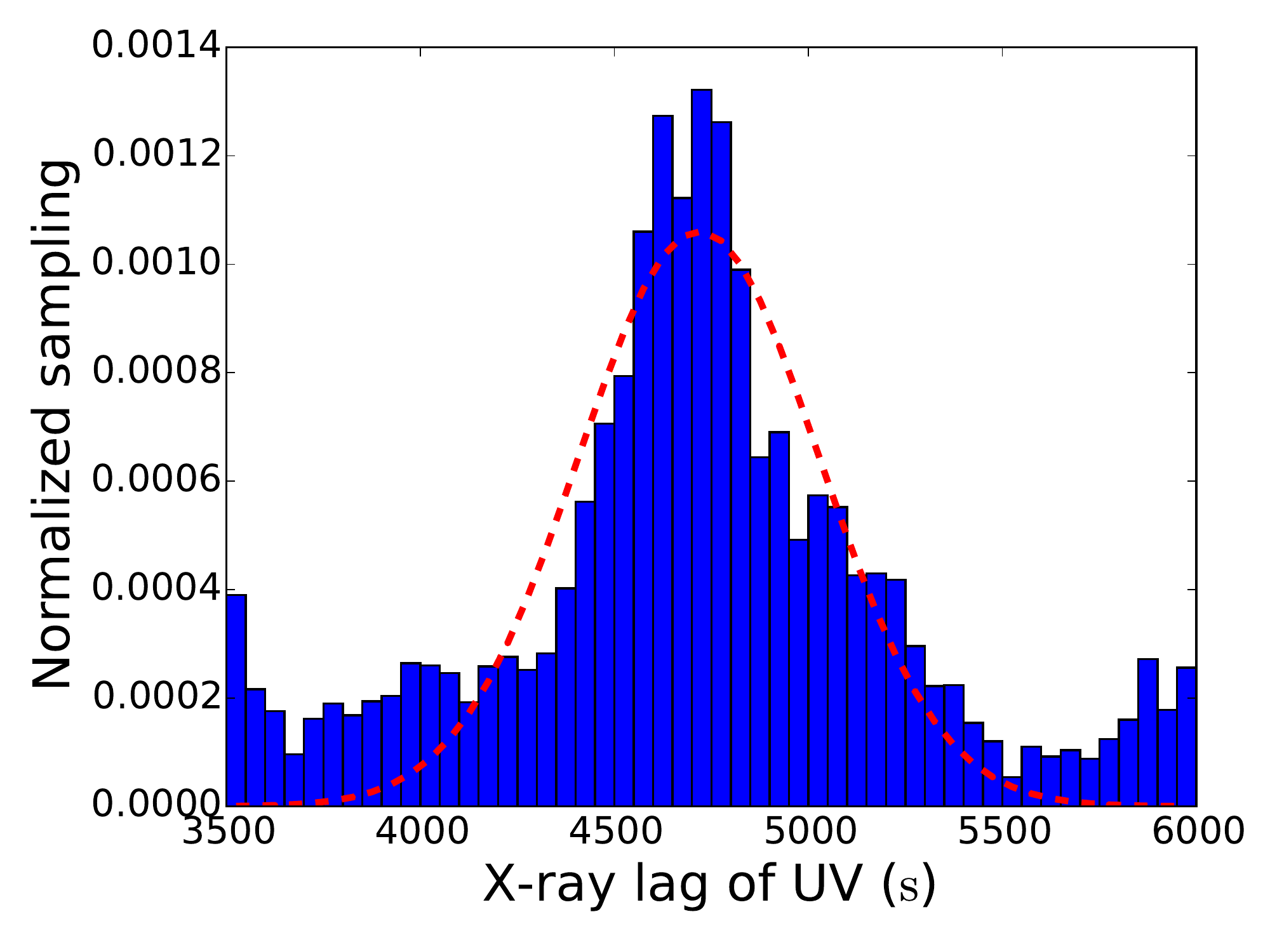}
\includegraphics[scale=0.21, angle=0]{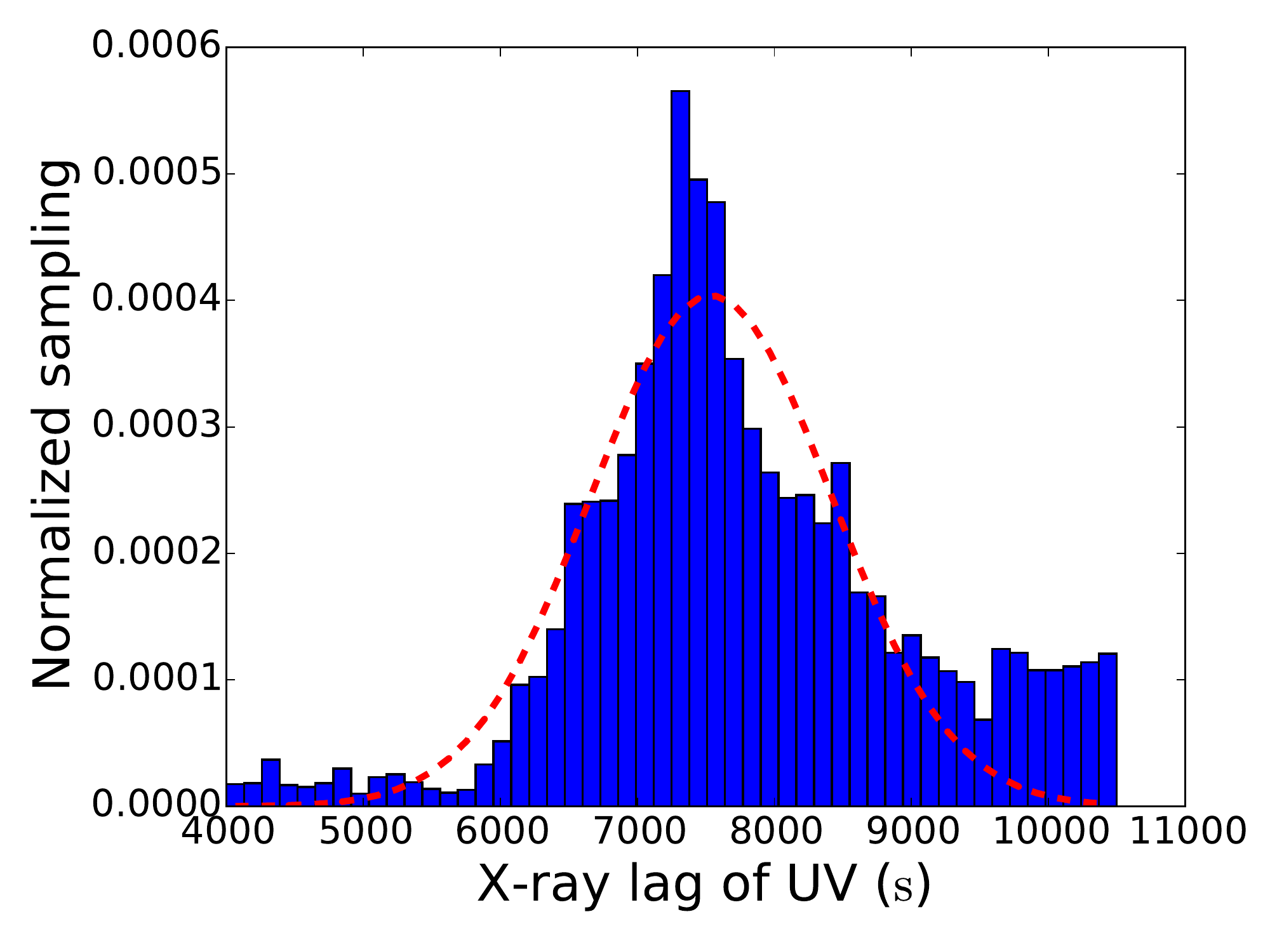}
\\
\end{tabular}
\caption{From left to right: JAVELIN cross-correlation plot between SX $and$ the UVW1 energy bands and HX $and$ UVW1 bands.}
\label{jav}
\end{figure}

To further validate our lag estimation, we employed the JAVELIN\footnote{https://bitbucket.org/nye17/javelin} code of \citet{2011ApJ...735...80Z}. Lags estimated from  this method are shown in Fig. \ref{jav}. By assuming a perfect Gaussian distribution of lags (depicted by the red dashed lines in the plots), we computed the mean and $1\sigma$ error on the lags to be $4.7\pm{0.3}\,\mathrm{ks}$, $7.5\pm{0.9}\,\mathrm{ks}$ and $4.7\pm{0.2}\,\mathrm{ks}$ respectively, for the UV--SX, UV--HX and the UV--X-ray (also shown in Table \ref{tab_ccf}).

\section{{Discussion \& Conclusion}}
We analyzed the $\sim100\,\mathrm{ks}$ simultaneous \textit{XMM-Newton} UV and X-ray data of the narrow-line Seyfert 1 galaxy \rm{Mrk~493} to search for possible correlated variability between these two bands. We found a significant correlation between these bands in which the UV emission lead the X-rays in their variability by $\sim5\,\mathrm{ks}$. We investigate the origin of the observed lags below.

\subsection{\textit{Accretion disk timescales}}
If the dominant emission from the accretion disc is a result of viscous heating in the disc, the photons emanating from different radii can be described as blackbodies with different temperatures \citep{2013peag.book.....N}. $\lambda_{eff}$ can be converted to the blackbody temperature which peaks in the particular band. Comparing this temperature to that of a standard accretion disc, we can calculate the disc radius and subsequently the light-crossing time between the X-ray source and the region of the disc wave-band with peak wavelength ($\lambda_{eff}$) as
\\
\begin{equation} 
t_{lc}\approx 2.6\times 10^{5}\Big(\frac{\lambda_{eff}}{3000\textup{\AA}}\Big)^{4/3}\Big(\frac{\dot{M}}{\dot{M}_{Edd}}\Big)^{1/3}\Big(\frac{M_{BH}}{10^{8}M_{\odot}}\Big)^{2/3}.\\
\end{equation}

The dynamical timescale corresponding to the Keplerian frequency can be written as
\\
\begin{equation} 
t_{dyn}=\Big(\frac{r^{3}}{GM_{BH}}\Big)^{1/2}\approx 500\Big(\frac{M_{BH}}{10^{8}M_{\odot}}\Big)\Big(\frac{r}{r_{g}}\Big)^{3/2}.\\
\end{equation}

The thermal timescale $t_{th}$ can be written assuming thermal equilibrium in the disc as follows
\\
\begin{equation} 
t_{th}=\frac{1}{\alpha}t_{dyn},\\
\end{equation}
and finally, since the radial inflow of matter is governed by viscosity, the viscous timescale $t_{vis}$ can be expressed as
\\
\begin{equation} 
t_{vis}\approx \frac{1}{\alpha}(\frac{r}{h})^{2}t_{dyn},\\
\end{equation}
where $r$ and $h$ are the radius and height of the disc respectively. $\alpha\sim0.1$ is the viscosity parameter and $r_{g}=GM_{BH}/c^{2}$

For \rm{Mrk~493} with $M_{BH}\sim1.5\times10^6M_{\odot}$ \citep{2014ApJ...793..108W} assuming moderately high Eddington scaled accretion rate of $0.1$, we calculated the various timescales associated with the accretion disc. We used $\lambda_{eff}$ of the UVW1 filter as $2910\,\mathrm{\textup{\AA}}$ \citep{2001A&A...365L..36M}.  
Our estimated values for the different timescales are $t_{lc}\sim7.0\,\mathrm{ks}$, $t_{dyn}\sim2.5\,\mathrm{days}$, $t_{th}\sim25\,\mathrm{days}$ and $t_{vis}\sim7\,\mathrm{years}$.

\subsection{\emph{Propagation fluctuation delay}}
The broadband variability properties of many accreting systems are usually explained in terms of inward propagating fluctuations in the accretion disc \citep[see e.g.,][]{
1997MNRAS.292..679L, 2006MNRAS.367..801A}.
In this model, the hot inner regions emitting soft X-rays and the outer cooler part of the disc responsible for the longer wavelength UV photons cannot exchange information faster than the sound crossing time $t_{sc}$. For accretion rate fluctuations propagating inward in the radial direction, the sound crossing time can be written as

\begin{equation}
t_{sc}=t_{dyn}\Big(\frac{r}{h}\Big),
\end{equation}
where $h$ is disc height and $r$ is the radius corresponding to the observed UV emission. We assumed $h/r\sim0.1$ \citep[see e.g.,][]{2006ASPC..360..265C} for this source, thus the fluctuation propagation timescale is of the order of $\sim2.2\times10^{6}\,\mathrm{sec}$. This is about two orders of magnitude longer than our measured UV/X-ray lag. Therefore, we rule out accretion rate fluctuation propagating inwards through the disc as being the possible origin of our measured UV/X-ray lag.

\subsection{\textit{Comptonization lag}}
Although, the light-crossing timescale provides reasonable explanation for the detected lag, the Comptonization process itself requires a finite time due to multiple Compton up-scattering. So the observed delay should be the combination of the light-crossing time $t_{lc}$ plus the time it will take the soft photons to be Comptonized in the hot electron plasma i.e. the Comptonization time $t_{comp}$ \citep[see e.g.,][]{1985ApJ...289..514Z,2006ApJ...651L..13D,2015ASInC..12...57D}.\\
A seed photon injected into a static Comptonizing corona with small optical depth $\tau$ ($\tau<<1$), electron temperature $kT_{e}$ increases its energy by a fraction 
\\
\begin{equation} 
A=1+4\theta+16\theta^{2},\\
\end{equation}

where $\theta=\frac{kT_{e}}{m_{e}c^{2}}$, $k$ is the Boltzmann's constant and $m_e$ is the mass of an electron. If the injected photon undergoes $n$ scatterings within the cloud before it escapes, its final energy is $E_{n}=A^{n}E_{0}$, where $E_{0}$ is the initial energy of the injected photon.

If the size of the X-ray emitting corona is $R_c$, then the photon mean free path $\lambda$ through the cloud can be expressed as $\lambda\sim\frac{R_{c}}{max(1,\tau)}$. Thus, the time interval between successive scatterings can be expressed as, $t_c=\frac{(R_c/c)}{max(1, \tau)}$.

Finally, the Comptonization time $t_{comp}$ required to upscatter a seed photon with energy $E_0$ to $E_n$ after $n$ scatterings will be;
\\
\begin{equation}
t_{comp}=nt_c.
\end{equation}
For Mrk 493, we considered a plausible scenario where UVW1 seed photons produced due to viscous heating in the disc are Compton upscattered into the observed X-rays after travelling the light-crossing time $t_{lc}$ to reach the compact corona of size $\sim20r_{g}$ \citep[see e.g.,][]{2013ApJ...769L...7R, 2017MNRAS.466.3951A}. We took the UV seed photon energy to be $\sim4.25\,\mathrm{eV}$ (for $\lambda_{eff}=2910\,\mathrm{\textup{\AA}}$) and therefore the time it will take to increase the energy of this photon to $\sim1.0\,\mathrm{keV}$ (the approximate soft X-ray midpoint energy) due to inverse Compton scattering in an electron cloud corona with $T_{e}\sim100\,\mathrm{keV}$ as calculated from the above relation will be $\sim900\,\mathrm{sec}$ while it will take $\sim1200\,\mathrm{sec}$ to increase the seed photon's energy to $5\,\mathrm{keV}$ (i.e. the approximate midpoint energy of the hard X-rays). This difference in Comptonization delay is expected as more scatterings will be required for the injected seed photon to be boosted to harder X-ray energies. Then, as stated above, the UV/X-ray time lag $t_{lag}$ should be;
\\
\begin{equation}
t_{lag}=t_{lc}+t_{comp}.
\end{equation}
Now, for $\dot{m}=0.1$ the expected lag from the above equations should be $\sim7.9\,\mathrm{ks}$ and $\sim8.2\,\mathrm{ks}$ for the UV-SX and UV-HX flux variabilities respectively. These values are in good agreement with our measured values within measurement uncertainties as shown in Table \ref{tab_ccf}. 
Even for a super-Eddington accretion rate with $\dot{m}\sim1.0$, 
the predicted value of lag will be $\sim16\,\mathrm{ks}$, only about a factor of 3 higher than our estimated lag.\\

A recent work on the \textit{XMM-Newton} X-ray data of \rm{Mrk~493} by \citet{2018MNRAS.tmp..812B} suggests the presence of strong reflection components and posit that the variations in X-rays are due to the changes in the degree of light bending in the vicinity of the central black hole. Therefore, one might expect the production of UV emission from thermal reprocessing of the illuminating X-rays in which the UV lag the X-rays. The absence of such a lag in our analysis implies that reprocessing probably does not play an important role in the UV/X-ray variability seen in this source. The most likely scenario being that due to strong light bending effect, strong coronal illumination is confined to the inner regions with no reprocessed emission in the UV band. This will be the case if the coronal height is considerably small (also suggested by \citet{2018MNRAS.tmp..812B}). Thus the observed delay can most plausibly be explained as Comptonization lag. 

We note that $\sim3\%$ variability
amplitude in the optical/UV band is unlikely 
to drive $\sim20\%$ variability amplitude in the X-rays by
thermal Comptonization alone in a static corona. The $0.3-10\,\mathrm{keV}$ band emission consists of the primary powerlaw arising from the thermal Comptonization, the soft X-ray excess and the reflection. The latter two components can introduce additional variability. If we filter out the rapid variability events in the $2-5\,\mathrm{keV}$ band lightcurve which is relatively free of soft excess and iron line, the variability amplitude becomes comparable to that of the UVW1 band. 
This shows that variations in the seed UV photons primarily drives the slower variability of the X-ray powerlaw
continuum. The soft X-ray
excess can arise either due to thermal Comptonization in a warm, optically thick material most likely the 
 innermost regions of the disc itself and/or blurred reflection. In the warm Comtonization scenorio, if the soft
excess and the UV emission are arising from the adjacent
regions \citep[see e.g.,][]{2018MNRAS.480.1247K}, the soft
X-ray excess can increase both due to increased seed
photons and increased energy dissipation in the warm
corona. Hence the soft band can probably vary strongly and still be correlated with the UV.\\

\acknowledgments

We thank Banibrata Mukhopadhyay, Iossif Papadakis and the anonymous referee for useful inputs that improved the manuscript. OA and MP respectively acknowledge support from IUCAA and the DSKPDF program of UGC, India.


\begin{thebibliography}{}
\makeatletter
\relax
\def\mn@urlcharsother{\let\do\@makeother \do\$\do\&\do\#\do\^\do\_\do\%\do\~}
\def\mn@doi{\begingroup\mn@urlcharsother \@ifnextchar [ {\mn@doi@}
  {\mn@doi@[]}}
\def\mn@doi@[#1]#2{\def\@tempa{#1}\ifx\@tempa\@empty \href
  {http://dx.doi.org/#2} {doi:#2}\else \href {http://dx.doi.org/#2} {#1}\fi
  \endgroup}
\def\mn@eprint#1#2{\mn@eprint@#1:#2::\@nil}
\def\mn@eprint@arXiv#1{\href {http://arxiv.org/abs/#1} {{\tt arXiv:#1}}}
\def\mn@eprint@dblp#1{\href {http://dblp.uni-trier.de/rec/bibtex/#1.xml}
  {dblp:#1}}
\def\mn@eprint@#1:#2:#3:#4\@nil{\def\@tempa {#1}\def\@tempb {#2}\def\@tempc
  {#3}\ifx \@tempc \@empty \let \@tempc \@tempb \let \@tempb \@tempa \fi \ifx
  \@tempb \@empty \def\@tempb {arXiv}\fi \@ifundefined
  {mn@eprint@\@tempb}{\@tempb:\@tempc}{\expandafter \expandafter \csname
  mn@eprint@\@tempb\endcsname \expandafter{\@tempc}}}

\bibitem[\protect\citeauthoryear{{Adegoke}, {Rakshit}  \&
  {Mukhopadhyay}}{{Adegoke} et~al.}{2017}]{2017MNRAS.466.3951A}
{Adegoke} O.,  {Rakshit} S.,   {Mukhopadhyay} B.,  2017, \mn@doi [\mnras]
  {10.1093/mnras/stw3320}, \href
  {http://adsabs.harvard.edu/abs/2017MNRAS.466.3951A} {466, 3951}

\bibitem[\protect\citeauthoryear{{Alexander}}{{Alexander}}{1997}]{1997ASSL..218..163A}
{Alexander} T.,  1997, in {Maoz} D.,  {Sternberg} A.,   {Leibowitz} E.~M.,
  eds,  Astrophysics and Space Science Library Vol. 218, Astronomical Time
  Series. p.~163, \mn@doi{10.1007/978-94-015-8941-3_14}

\bibitem[\protect\citeauthoryear{{Alexander}}{{Alexander}}{2013}]{2013arXiv1302.1508A}
{Alexander} T.,  2013, preprint, \href
  {http://adsabs.harvard.edu/abs/2013arXiv1302.1508A} {} (\mn@eprint {arXiv}
  {1302.1508})


\bibitem[\protect\citeauthoryear{{Ar{\'e}valo} \& {Uttley}}{{Ar{\'e}valo} \&
  {Uttley}}{2006}]{2006MNRAS.367..801A}
{Ar{\'e}valo} P.,  {Uttley} P.,  2006, \mn@doi [\mnras]
  {10.1111/j.1365-2966.2006.09989.x}, \href
  {http://adsabs.harvard.edu/abs/2006MNRAS.367..801A} {367, 801}

\bibitem[\protect\citeauthoryear{{Ar{\'e}valo}, {Papadakis}, {Kuhlbrodt}  \&
  {Brinkmann}}{{Ar{\'e}valo} et~al.}{2005}]{2005A&A...430..435A}
{Ar{\'e}valo} P.,  {Papadakis} I.,  {Kuhlbrodt} B.,   {Brinkmann} W.,  2005,
  \mn@doi [\aap] {10.1051/0004-6361:20041801}, \href
  {http://adsabs.harvard.edu/abs/2005A%26A...430..435A} {430, 435}

\bibitem[\protect\citeauthoryear{{Ar{\'e}valo}, {Uttley}, {Kaspi}, {Breedt},
  {Lira}  \& {McHardy}}{{Ar{\'e}valo} et~al.}{2008}]{2008MNRAS.389.1479A}
{Ar{\'e}valo} P.,  {Uttley} P.,  {Kaspi} S.,  {Breedt} E.,  {Lira} P.,
  {McHardy} I.~M.,  2008, \mn@doi [\mnras] {10.1111/j.1365-2966.2008.13719.x},
  \href {http://adsabs.harvard.edu/abs/2008MNRAS.389.1479A} {389, 1479}

\bibitem[\protect\citeauthoryear{{Ar{\'e}valo}, {Uttley}, {Lira}, {Breedt},
  {McHardy}  \& {Churazov}}{{Ar{\'e}valo} et~al.}{2009}]{2009MNRAS.397.2004A}
{Ar{\'e}valo} P.,  {Uttley} P.,  {Lira} P.,  {Breedt} E.,  {McHardy} I.~M.,
  {Churazov} E.,  2009, \mn@doi [\mnras] {10.1111/j.1365-2966.2009.15110.x},
  \href {http://adsabs.harvard.edu/abs/2009MNRAS.397.2004A} {397, 2004}

\bibitem[\protect\citeauthoryear{{Bonson}, {Gallo}, {Wilkins}  \&
  {Fabian}}{{Bonson} et~al.}{2018}]{2018MNRAS.tmp..812B}
{Bonson} K.,  {Gallo} L.~C.,  {Wilkins} D.~R.,   {Fabian} A.~C.,  2018, \mn@doi
  [\mnras] {10.1093/mnras/sty828}, \href
  {http://adsabs.harvard.edu/abs/2018MNRAS.tmp..812B} {}

\bibitem[\protect\citeauthoryear{{Breedt} et~al.,}{{Breedt}
  et~al.}{2009}]{2009MNRAS.394..427B}
{Breedt} E.,  et~al., 2009, \mn@doi [\mnras]
  {10.1111/j.1365-2966.2008.14302.x}, \href
  {http://adsabs.harvard.edu/abs/2009MNRAS.394..427B} {394, 427}

\bibitem[\protect\citeauthoryear{{Breedt} et~al.,}{{Breedt}
  et~al.}{2010}]{2010MNRAS.403..605B}
{Breedt} E.,  et~al., 2010, \mn@doi [\mnras]
  {10.1111/j.1365-2966.2009.16146.x}, \href
  {http://adsabs.harvard.edu/abs/2010MNRAS.403..605B} {403, 605}

\bibitem[\protect\citeauthoryear{{Buisson} et~al.,}{{Buisson}
  et~al.}{2018}]{2018MNRAS.475.2306B}
{Buisson} D.~J.~K.,  et~al., 2018, \mn@doi [\mnras] {10.1093/mnras/sty008},
  \href {http://adsabs.harvard.edu/abs/2018MNRAS.475.2306B} {475, 2306}

\bibitem[\protect\citeauthoryear{{Cackett}, {Horne}  \& {Winkler}}{{Cackett}
  et~al.}{2007}]{2007MNRAS.380..669C}
{Cackett} E.~M.,  {Horne} K.,   {Winkler} H.,  2007, \mn@doi [\mnras]
  {10.1111/j.1365-2966.2007.12098.x}, \href
  {http://adsabs.harvard.edu/abs/2007MNRAS.380..669C} {380, 669}

\bibitem[\protect\citeauthoryear{{Cameron}, {McHardy}, {Dwelly}, {Breedt},
  {Uttley}, {Lira}  \& {Arevalo}}{{Cameron} et~al.}{2012}]{2012MNRAS.422..902C}
{Cameron} D.~T.,  {McHardy} I.,  {Dwelly} T.,  {Breedt} E.,  {Uttley} P.,
  {Lira} P.,   {Arevalo} P.,  2012, \mn@doi [\mnras]
  {10.1111/j.1365-2966.2012.20677.x}, \href
  {http://adsabs.harvard.edu/abs/2012MNRAS.422..902C} {422, 902}


\bibitem[\protect\citeauthoryear{{Czerny}}{{Czerny}}{2006}]{2006ASPC..360..265C}
{Czerny} B.,  2006, in {Gaskell} C.~M.,  {McHardy} I.~M.,  {Peterson} B.~M.,
  {Sergeev} S.~G.,  eds,  Astronomical Society of the Pacific Conference Series
  Vol. 360, Astronomical Society of the Pacific Conference Series. p.~265

\bibitem[\protect\citeauthoryear{{Dasgupta} \& {Rao}}{{Dasgupta} \&
  {Rao}}{2006}]{2006ApJ...651L..13D}
{Dasgupta} S.,  {Rao} A.~R.,  2006, \mn@doi [\apjl] {10.1086/509117}, \href
  {http://adsabs.harvard.edu/abs/2006ApJ...651L..13D} {651, L13}

\bibitem[\protect\citeauthoryear{{Dewangan}, {Pawar}  \& {Pal}}{{Dewangan}
  et~al.}{2015}]{2015ASInC..12...57D}
{Dewangan} G.~C.,  {Pawar} P.~K.,   {Pal} M.,  2015, in Astronomical Society of
  India Conference Series.

\bibitem[\protect\citeauthoryear{{Edelson} \& {Krolik}}{{Edelson} \&
  {Krolik}}{1988}]{1988ApJ...333..646E}
{Edelson} R.~A.,  {Krolik} J.~H.,  1988, \mn@doi [\apj] {10.1086/166773}, \href
  {http://adsabs.harvard.edu/abs/1988ApJ...333..646E} {333, 646}

\bibitem[\protect\citeauthoryear{{Edelson} et~al.,}{{Edelson}
  et~al.}{2015}]{2015ApJ...806..129E}
{Edelson} R.,  et~al., 2015, \mn@doi [\apj] {10.1088/0004-637X/806/1/129},
  \href {http://adsabs.harvard.edu/abs/2015ApJ...806..129E} {806, 129}

\bibitem[\protect\citeauthoryear{{Edelson} et~al.,}{{Edelson}
  et~al.}{2017}]{2017ApJ...840...41E}
{Edelson} R.,  et~al., 2017, \mn@doi [\apj] {10.3847/1538-4357/aa6890}, \href
  {http://adsabs.harvard.edu/abs/2017ApJ...840...41E} {840, 41}

\bibitem[\protect\citeauthoryear{{Fabian}, {Lohfink}, {Kara}, {Parker},
  {Vasudevan}  \& {Reynolds}}{{Fabian} et~al.}{2015}]{2015MNRAS.451.4375F}
{Fabian} A.~C.,  {Lohfink} A.,  {Kara} E.,  {Parker} M.~L.,  {Vasudevan} R.,
  {Reynolds} C.~S.,  2015, \mn@doi [\mnras] {10.1093/mnras/stv1218}, \href
  {http://adsabs.harvard.edu/abs/2015MNRAS.451.4375F} {451, 4375}

\bibitem[\protect\citeauthoryear{{Gardner} \& {Done}}{{Gardner} \&
  {Done}}{2017}]{2017MNRAS.470.3591G}
{Gardner} E.,  {Done} C.,  2017, \mn@doi [\mnras] {10.1093/mnras/stx946}, \href
  {http://adsabs.harvard.edu/abs/2017MNRAS.470.3591G} {470, 3591}

\bibitem[\protect\citeauthoryear{{Gaskell}}{{Gaskell}}{2007}]{2007ASPC..373..596G}
{Gaskell} C.~M.,  2007, in {Ho} L.~C.,  {Wang} J.-W.,  eds,  Astronomical
  Society of the Pacific Conference Series Vol. 373, The Central Engine of
  Active Galactic Nuclei. p.~596 (\mn@eprint {} {astro-ph/0612474})

\bibitem[\protect\citeauthoryear{{Gliozzi}, {Papadakis}, {Grupe}, {Brinkmann}
  \& {R{\"a}th}}{{Gliozzi} et~al.}{2013}]{2013MNRAS.433.1709G}
{Gliozzi} M.,  {Papadakis} I.~E.,  {Grupe} D.,  {Brinkmann} W.~P.,   {R{\"a}th}
  C.,  2013, \mn@doi [\mnras] {10.1093/mnras/stt848}, \href
  {http://adsabs.harvard.edu/abs/2013MNRAS.433.1709G} {433, 1709}

\bibitem[\protect\citeauthoryear{{Haardt} \& {Maraschi}}{{Haardt} \&
  {Maraschi}}{1993}]{1993ApJ...413..507H}
{Haardt} F.,  {Maraschi} L.,  1993, \mn@doi [\apj] {10.1086/173020}, \href
  {http://adsabs.harvard.edu/abs/1993ApJ...413..507H} {413, 507}

\bibitem[\protect\citeauthoryear{{Jansen} et~al.,}{{Jansen}
  et~al.}{2001}]{2001A&A...365L...1J}
{Jansen} F.,  et~al., 2001, \mn@doi [\aap] {10.1051/0004-6361:20000036}, \href
  {http://adsabs.harvard.edu/abs/2001A%26A...365L...1J} {365, L1}


\bibitem[\protect\citeauthoryear{{Koratkar} \& {Blaes}}{{Koratkar} \&
  {Blaes}}{1999}]{1999PASP..111....1K}
{Koratkar} A.,  {Blaes} O.,  1999, \mn@doi [\pasp] {10.1086/316294}, \href
  {http://adsabs.harvard.edu/abs/1999PASP..111....1K} {111, 1}

\bibitem[\protect\citeauthoryear{{Krolik}, {Horne}, {Kallman}, {Malkan},
  {Edelson}  \& {Kriss}}{{Krolik} et~al.}{1991}]{1991ApJ...371..541K}
{Krolik} J.~H.,  {Horne} K.,  {Kallman} T.~R.,  {Malkan} M.~A.,  {Edelson}
  R.~A.,   {Kriss} G.~A.,  1991, \mn@doi [\apj] {10.1086/169918}, \href
  {http://adsabs.harvard.edu/abs/1991ApJ...371..541K} {371, 541}

\bibitem[\protect\citeauthoryear{{Kubota} \& {Done}}{{Kubota} \&
  {Done}}{2018}]{2018MNRAS.480.1247K}
{Kubota} A.,  {Done} C.,  2018, \mn@doi [\mnras] {10.1093/mnras/sty1890}, \href
  {http://adsabs.harvard.edu/abs/2018MNRAS.480.1247K} {480, 1247}

\bibitem[\protect\citeauthoryear{{Lyubarskii}}{{Lyubarskii}}{1997}]{1997MNRAS.292..679L}
{Lyubarskii} Y.~E.,  1997, \mn@doi [\mnras] {10.1093/mnras/292.3.679}, \href
  {http://adsabs.harvard.edu/abs/1997MNRAS.292..679L} {292, 679}

\bibitem[\protect\citeauthoryear{{Maoz}, {Markowitz}, {Edelson}  \&
  {Nandra}}{{Maoz} et~al.}{2002}]{2002AJ....124.1988M}
{Maoz} D.,  {Markowitz} A.,  {Edelson} R.,   {Nandra} K.,  2002, \mn@doi [\aj]
  {10.1086/342937}, \href {http://adsabs.harvard.edu/abs/2002AJ....124.1988M}
  {124, 1988}

\bibitem[\protect\citeauthoryear{{Markowitz} et~al.,}{{Markowitz}
  et~al.}{2003}]{2003ApJ...593...96M}
{Markowitz} A.,  et~al., 2003, \mn@doi [\apj] {10.1086/375330}, \href
  {http://adsabs.harvard.edu/abs/2003ApJ...593...96M} {593, 96}


\bibitem[\protect\citeauthoryear{{Mason} et~al.,}{{Mason}
  et~al.}{2001}]{2001A&A...365L..36M}
{Mason} K.~O.,  et~al., 2001, \mn@doi [\aap] {10.1051/0004-6361:20000044},
  \href {http://adsabs.harvard.edu/abs/2001A%26A...365L..36M} {365, L36}


\bibitem[\protect\citeauthoryear{{McHardy} et~al.,}{{McHardy}
  et~al.}{2018}]{2018MNRAS.tmp.1900M}
{McHardy} I.~M.,  et~al., 2018, \mn@doi [\mnras] {10.1093/mnras/sty1983}, \href
  {http://adsabs.harvard.edu/abs/2018MNRAS.tmp.1900M} {}


\bibitem[\protect\citeauthoryear{{Nandra}}{{Nandra}}{2001}]{2001AdSpR..28..295N}
{Nandra} K.,  2001, \mn@doi [Advances in Space Research]
  {10.1016/S0273-1177(01)00409-4}, \href
  {http://adsabs.harvard.edu/abs/2001AdSpR..28..295N} {28, 295}

\bibitem[\protect\citeauthoryear{{Netzer}}{{Netzer}}{2013}]{2013peag.book.....N}
{Netzer} H.,  2013, {The Physics and Evolution of Active Galactic Nuclei}

\bibitem[\protect\citeauthoryear{{Osterbrock} \& {Pogge}}{{Osterbrock} \&
  {Pogge}}{1985}]{1985ApJ...297..166O}
{Osterbrock} D.~E.,  {Pogge} R.~W.,  1985, \mn@doi [\apj] {10.1086/163513},
  \href {http://adsabs.harvard.edu/abs/1985ApJ...297..166O} {297, 166}

\bibitem[\protect\citeauthoryear{{Pal}, {Dewangan}, {Connolly}  \&
  {Misra}}{{Pal} et~al.}{2017}]{2017MNRAS.466.1777P}
{Pal} M.,  {Dewangan} G.~C.,  {Connolly} S.~D.,   {Misra} R.,  2017, \mn@doi
  [\mnras] {10.1093/mnras/stw3173}, \href
  {http://adsabs.harvard.edu/abs/2017MNRAS.466.1777P} {466, 1777}

\bibitem[\protect\citeauthoryear{{Pawar}, {Dewangan}, {Papadakis}, {Patil},
  {Pal}  \& {Kembhavi}}{{Pawar} et~al.}{2017}]{2017MNRAS.472.2823P}
{Pawar} P.~K.,  {Dewangan} G.~C.,  {Papadakis} I.~E.,  {Patil} M.~K.,  {Pal}
  M.,   {Kembhavi} A.~K.,  2017, \mn@doi [\mnras] {10.1093/mnras/stx2163},
  \href {http://adsabs.harvard.edu/abs/2017MNRAS.472.2823P} {472, 2823}

\bibitem[\protect\citeauthoryear{{Peterson}, {Wanders}, {Horne}, {Collier},
  {Alexander}, {Kaspi}  \& {Maoz}}{{Peterson}
  et~al.}{1998}]{1998PASP..110..660P}
{Peterson} B.~M.,  {Wanders} I.,  {Horne} K.,  {Collier} S.,  {Alexander} T.,
  {Kaspi} S.,   {Maoz} D.,  1998, \mn@doi [\pasp] {10.1086/316177}, \href
  {http://adsabs.harvard.edu/abs/1998PASP..110..660P} {110, 660}

\bibitem[\protect\citeauthoryear{{Reis} \& {Miller}}{{Reis} \&
  {Miller}}{2013}]{2013ApJ...769L...7R}
{Reis} R.~C.,  {Miller} J.~M.,  2013, \mn@doi [\apjl]
  {10.1088/2041-8205/769/1/L7}, \href
  {http://adsabs.harvard.edu/abs/2013ApJ...769L...7R} {769, L7}


\bibitem[\protect\citeauthoryear{{Shemmer}, {Uttley}, {Netzer}  \&
  {McHardy}}{{Shemmer} et~al.}{2003}]{2003MNRAS.343.1341S}
{Shemmer} O.,  {Uttley} P.,  {Netzer} H.,   {McHardy} I.~M.,  2003, \mn@doi
  [\mnras] {10.1046/j.1365-8711.2003.06775.x}, \href
  {http://adsabs.harvard.edu/abs/2003MNRAS.343.1341S} {343, 1341}

\bibitem[\protect\citeauthoryear{{Str{\"u}der} et~al.,}{{Str{\"u}der}
  et~al.}{2001}]{2001A&A...365L..18S}
{Str{\"u}der} L.,  et~al., 2001, \mn@doi [\aap] {10.1051/0004-6361:20000066},
  \href {http://adsabs.harvard.edu/abs/2001A%26A...365L..18S} {365, L18}


\bibitem[\protect\citeauthoryear{{Troyer}, {Starkey}, {Cackett}, {Bentz},
  {Goad}, {Horne}  \& {Seals}}{{Troyer} et~al.}{2016}]{2016MNRAS.456.4040T}
{Troyer} J.,  {Starkey} D.,  {Cackett} E.~M.,  {Bentz} M.~C.,  {Goad} M.~R.,
  {Horne} K.,   {Seals} J.~E.,  2016, \mn@doi [\mnras] {10.1093/mnras/stv2862},
  \href {http://adsabs.harvard.edu/abs/2016MNRAS.456.4040T} {456, 4040}

\bibitem[\protect\citeauthoryear{{Turner} et~al.,}{{Turner}
  et~al.}{2001}]{2001A&A...365L..27T}
{Turner} M.~J.~L.,  et~al., 2001, \mn@doi [\aap] {10.1051/0004-6361:20000087},
  \href {http://adsabs.harvard.edu/abs/2001A%26A...365L..27T} {365, L27}

\bibitem[\protect\citeauthoryear{{Uttley} \& {Mchardy}}{{Uttley} \&
  {Mchardy}}{2004}]{2004PThPS.155..170U}
{Uttley} P.,  {Mchardy} I.~M.,  2004, \mn@doi [Progress of Theoretical Physics
  Supplement] {10.1143/PTPS.155.170}, \href
  {http://adsabs.harvard.edu/abs/2004PThPS.155..170U} {155, 170}

\bibitem[\protect\citeauthoryear{{Vaughan}, {Edelson}, {Warwick}  \&
  {Uttley}}{{Vaughan} et~al.}{2003}]{2003MNRAS.345.1271V}
{Vaughan} S.,  {Edelson} R.,  {Warwick} R.~S.,   {Uttley} P.,  2003, \mn@doi
  [\mnras] {10.1046/j.1365-2966.2003.07042.x}, \href
  {http://adsabs.harvard.edu/abs/2003MNRAS.345.1271V} {345, 1271}


\bibitem[\protect\citeauthoryear{{Wang} et~al.,}{{Wang}
  et~al.}{2014}]{2014ApJ...793..108W}
{Wang} J.-M.,  et~al., 2014, \mn@doi [\apj] {10.1088/0004-637X/793/2/108},
  \href {http://adsabs.harvard.edu/abs/2014ApJ...793..108W} {793, 108}

\bibitem[\protect\citeauthoryear{{Zdziarski}}{{Zdziarski}}{1985}]{1985ApJ...289..514Z}
{Zdziarski} A.~A.,  1985, \mn@doi [\apj] {10.1086/162912}, \href
  {http://adsabs.harvard.edu/abs/1985ApJ...289..514Z} {289, 514}

\bibitem[\protect\citeauthoryear{{Zu}, {Kochanek}  \& {Peterson}}{{Zu}
  et~al.}{2011}]{2011ApJ...735...80Z}
{Zu} Y.,  {Kochanek} C.~S.,   {Peterson} B.~M.,  2011, \mn@doi [\apj]
  {10.1088/0004-637X/735/2/80}, \href
  {http://adsabs.harvard.edu/abs/2011ApJ...735...80Z} {735, 80}


\makeatother
\end{thebibliography}
\end{document}